\documentstyle[aps,multicol,prl,epsfig]{revtex}
\def\beq{\begin{equation}}
\def\eeq{\end{equation}}
\def\be{\begin{eqnarray}}
\def\ee{\end{eqnarray}}
\def\ci{\cite}
\def\bi{\bibitem}

\def\PkE{P({\bf k},E)}

\begin{document}

\draft 
\twocolumn[\hsize\textwidth\columnwidth\hsize\csname@twocolumnfalse%
\endcsname

\title{Interpretation of $y$-scaling of the nuclear response}

\author{Omar Benhar}

\address {INFN, Sezione Roma 1 \\ 
 Dipartimento di Fisica, Universit\`a ``La Sapienza" \\
Piazzale Aldo Moro, 2. I-00185 Roma, Italy\\ }

\date{\today}
\maketitle


\begin{abstract}

The behavior of the nuclear matter response in the region of
large momentum transfer, in which plane wave impulse
approximation predicts the onset of $y$-scaling, is discussed.
The theoretical analysis shows that scaling violations produced
by final state interactions are driven by the
momentum dependence of the nucleon-nucleon scattering cross section. 
 Their study may provide valuable information on possible
modifications of nucleon-nucleon scattering in the nuclear medium.  

\medskip

\end{abstract}


]


Inclusive scattering of high energy electrons off nuclear targets has
long been recognized as a powerful tool to measure the nucleon
momentum and removal energy distribution \cite{cg}. The underlying 
picture is that 
at large momentum transfer electron-nucleus scattering reduces
to the incoherent sum of elementary scattering processes
involving individual nucleons, distributed in momentum and removal energy
according to the spectral function $\PkE$. In the early seventies, West
first pointed out \ci{West} that, if the electron-nucleon processes are
elastic and 
the final state state interactions between the struck particle and the 
spectator system can be neglected, the nuclear response $S({\bf q},\omega)$, 
which generally depends upon both momentum (${\bf q}$) and energy ($\omega$)
transfer, exhibit scaling, i.e. it can be simply related to a function of
only one kinematical variable, denoted $y$. Within the simplest nonrelativistic
approximation, $y$ can be identified with the minimum projection of the nucleon 
momentum along the direction of the momentum transfer, while the scaling function
$F(y) = (q/m)S({\bf q},\omega)$, where $m$ denotes the nucleon mass, 
can be directly written in terms of the nucleon momentum distribution.

The approach to $y$-scaling in few-nucleon systems
\cite{3hescal,2hscal} and medium-heavy nuclei \cite{NE3} has been 
experimentally investigated
at SLAC in the kinematical domain extending up to
$Q^2  = q^2 - \omega^2 \sim$ 3 (GeV/c)$^2$. Recent measurements
carried out at TJNAF using Carbon, Iron and Gold targets have extended
the $Q^2$ range up to $\sim$ 7 (GeV/c)$^2$ \ci{TJNAF}.
The available data, plotted as a function of $y$, display a striking 
overall scaling behavior at $y < 0$, corresponding to 
$\omega < \omega_{QE}$ ($\omega_{QE} =  Q^2/2m$, is the energy transfer
associated with elastic scattering off a free nucleon at rest), indicating 
that elastic scattering off individual nucleons is indeed the dominant 
reaction mechanism in that region. However, the analysis of the $Q^2$-dependence 
at fixed $y$ shows sizeable scaling violations at $y < $ $-0.2$ GeV/c and 
$Q^2 <$ 3 (GeV/c)$^2$, to be ascribed mainly to final state interactions (FSI).
Recenlty, Donnelly and Sick \ci{super} have also shown that the $y$-scaling
functions of different nuclei exhibit scaling of the {\it second kind} 
in the new variable $\psi^\prime = y/k_F$, $k_F$ being
the Fermi momentum.

The relevance of FSI in the kinematical regime of the SLAC
data has been systematically studied in refs.\ci{gangofsix,bp,LDAnp}, 
within a microscopic many-body approach in which nucleon-nucleon 
 (NN) correlations are consistently taken into account in both the initial 
and final state.
The results of these calculations confirm that FSI 
effects are important, particularly in the region of large negative $y$,
and have to be included to quantitatively account for the measured cross 
sections. 

The extension of the analysis of refs.\ci{gangofsix,bp,LDAnp} 
to the region 3~$< Q^2 <$~7~(GeV/c)$^2$, relevant to the 
interpretation of the new TJNAF data, requires the full 
calculation of the 
nuclear cross section, including the electromagnetic vertex, and the 
use of spectral functions adequate to describe finite targets, 
that can be obtained within the local density approximation \ci{LDAnp}.
In this paper we avoid these complications and focus only on the mechanisms 
driving the approach to $y$-scaling of $S({\bf q},\omega)$ in the case of 
infinite nuclear matter, a system whose spectral function can be obtained
from {\it ab initio} microscopic calculations using realistic nuclear 
hamiltonians \ci{bff}.

The most popular argument supporting the expectation that FSI effects become 
negligibly 
small at large momentum transfer is based on the observation that,  
compared to the PWIA amplitude, the amplitude of the process including a rescattering 
in the final state involves an extra propagator, associated with the
struck nucleon carrying a momentum $\sim {\bf q}$. As a consequence, 
this amplitude is expected to be suppressed when $q$ is large.
The validity of this argument has been proved in the context of a fully 
nonrelativistic model based on G-matrix perturbation theory, in which FSI were
described in terms of elastic NN scattering processes \ci{butler}. Within 
this picture, the disappearance of FSI effects at 
large momentum transfer simply reflects the fact that the elastic NN scattering 
cross section is a rapidly decreasing function of $q$. However, 
when the momentum of the struck nucleon is in the 2--5 GeV/c 
range, typical of the TJNAF kinematical regime, inelastic scattering is known 
to become dominant 
and must be taken into account. The inelastic cross section is roughly
momentum independent, leading to a scattering amplitude that grows linearly with $q$.
Therefore, in spite of the suppression coming from the nucleon 
propagators, inelastic rescatterings may give rise to FSI effects that remain nearly 
constant as the momentum transfer increases, thus preventing the onset of the 
$y$-scaling regime.

The treatment of FSI developed in ref.\ci{gangofsix} is based on the 
high-energy approximation, i.e. on the assumptions that i) the 
fast struck nucleon moves along a straight trajectory with 
momentum $\sim {\bf q}$ (eikonal approximation) and ii) the spectator 
system can be seen as a collection of fixed scattering centers 
(frozen approximation). The resulting quasielastic response reads 
\beq
S({\bf q},\omega) = \int d\omega^\prime S_{PWIA}({\bf q},\omega^\prime)
F_q(\omega - \omega^\prime)\ ,
\label{response}
\eeq
where the PWIA response is given by
\begin{eqnarray}
\nonumber
S_{PWIA}({\bf q},\omega) & = &  \int d^3k~dE~P({\bf k},E)~ \\
 & \times & \delta(\omega - E - \sqrt{|{\bf k}+{\bf q}|^2 + m^2} + m)\ ,
\label{PWIA:resp}
\end{eqnarray}
and the folding function 
$F_q(\omega)$ is defined as
\beq
F_q(\omega) = \int_{-\infty}^{+\infty} \frac{dt}{2\pi}\ {\rm e}^{i\omega t}
\ U_q(t)\ .
\label{folding}
\eeq

The effects of FSI are described by the function $U_q(t)$, which can 
be written \ci{bho}
\beq
U_q(t) = \frac{1}{A} \left\langle\ \sum_{i=1}^{A}
U_q^{(i)}(R,t) \right\rangle\ ,
\label{def:U}
\eeq
where $\langle \ldots \rangle$ denotes the expectation value in the target
ground state, $R\equiv\{ {\bf r}_1,{\bf r}_2,..,{\bf r}_A \}$
specifies the target configuration and 
\begin{eqnarray}
\nonumber 
U_q^{(i)}(R,z) & = & 1 + \left(\frac{i}{q}\right) \int_0^z dz_1
\sum_{j \neq i} \Gamma_q(| {\bf r}_{i} + {\hat q}z_1 - {\bf r}_{j} |) \\
\nonumber
  & + & \frac{1}{2}\left(\frac{i}{q}\right)^2 
\int_0^z dz_1 \sum_{j \neq i} 
\Gamma_q(| {\bf r}_{i} + {\hat q}z_1 - {\bf r}_{j} |) \\
  & \times & \int_0^z dz_2 \sum_{k \neq j \neq i}
\Gamma_q(|{\bf r}_{i} + {\hat q}z_2 - {\bf r}_{k} |) + \ldots\ .  
\label{expansion}
\end{eqnarray}
In the above equation ${\hat q} = ({\bf q}/q)$, while $z = vt$ is the 
distance travelled by the struck particle
during a time $t$ past the electromagnetic interaction.  
The dynamics of the rescattering process is dictated by the function 
$\Gamma_q(r)$, simply related to the amplitude for NN scattering
at incident momentum $q$ and momentum transfer ${\bf k}^\prime$, 
 $f_q({\bf k}^\prime)$, through
\beq
\Gamma_q(r) = \int \frac{d^3 k^\prime}{(2\pi)^2}\ 
f_q({\bf k}^\prime)\ {\rm e}^{i{\bf k}^\prime \cdot {\bf r}}\ .
\label{NN:ampl}
\eeq

Eqs.(\ref{response})-(\ref{NN:ampl}) clearly show 
that in absence of FSI $U_q(t) \equiv 1$, implying in turn 
$F_q(\omega) = \delta(\omega)$, and the PWIA result is recovered.
From eq.(\ref{expansion}) it is also apparent that, as expected, 
processes involving $n$ rescatterings exhibit a $(1/q)^n$ 
dependence associated with the nucleon propagators.
However, the presence of the $(1/q)^n$ 
factors does not guarantee that FSI corrections become vanishingly small 
at large $q$, since $U_q^{(i)}(R,z)$ has an additional $q$-dependence 
coming from the scattering amplitude $f_q({\bf k}^\prime)$. 

According to the optical theorem, 
the imaginary part of the forward amplitude, which is known to be dominant at 
high $q$, can be written in terms of the total scattering cross section 
$\sigma_{tot}(q)$ as
\beq
{\it Im}\ f_q(0) = \frac{q}{4\pi}\ \sigma_{tot}(q)\ .
\label{optical}
\eeq

Eqs.(\ref{expansion})-(\ref{optical}) show that the $(1/q)^n$ factors
appearing in the $n$-th rescattering contribution to $U_q^{(i)}(R,z)$
cancel, and the $q$-dependence of the FSI corrections is eventually driven by 
$\sigma_{tot}(q)$. 

\begin{figure}
\centerline
{\epsfig{figure=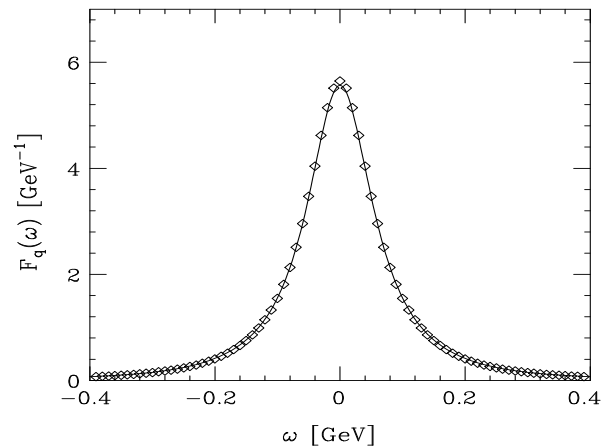,angle=090,width=9.75cm,height=6.1cm}}
\caption{
Folding functions in infinite nuclear matter, evaluated using 
eqs.(\protect\ref{folding})-(\protect\ref{NN:ampl}) and the parametrization
of the NN scattering amplitude of refs.\protect\ci{NNAMP1,NNAMP2}. 
The solid line and the diamonds correspond to $q$ = 2.2 and 3.4 GeV/c,
respectively. 
}
\label{ff}
\end{figure}

The measured NN total cross section is roughly constant 
in the momentum range 2--5~GeV/c \ci{NN:data}. Hence, in this region 
the $F_q(\omega)$ obtained from eqs.(\ref{folding})-(\ref{expansion})
using the free-space $\sigma_{tot}(q)$ does not show any appreciable
$q$-dependence. This feature is illustrated in fig. \ref{ff}, where 
the folding functions evaluated in infinite nuclear matter at $q =$ 2.2 and
3.4 GeV/c are compared.

\begin{figure}
\centerline{\epsfig{figure=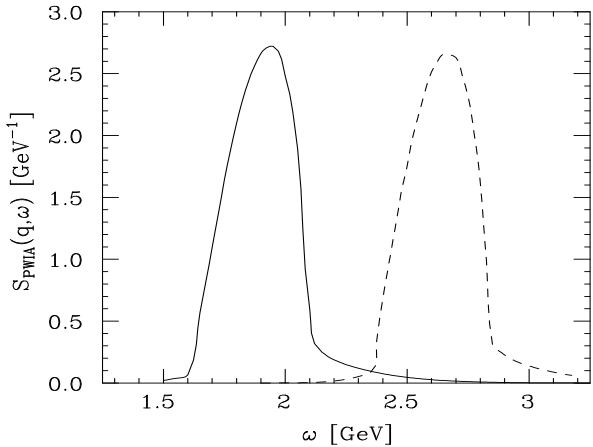,angle=090,width=9.50cm,height=6.1cm}}
\caption{
Nuclear matter response functions evaluated within PWIA using 
eq.(\protect\ref{PWIA:resp}) and the spectral function of ref.\protect\ci{bff}.
The solid and dashed lines correspond to $q$ = 2.6 and 3.4 GeV/c,
respectively.
}
\label{resp}
\end{figure}

At large $q$, the shape of the PWIA response, defined as in 
eq.(\ref{PWIA:resp}), also becomes nearly independent of $q$,
 as shown in fig. \ref{resp}. The nuclear matter $S_{PWIA}({\bf q},\omega)$ 
exhibits a bump whose maximum, corresponding to
$\omega \sim \omega_{QE} = \sqrt{q^2 + m^2} - m$, moves towards
higher values of $\omega$ as $q$ increases, whereas its height and width 
remain almost
constant, being dictated by the Fermi momentum $k_F$. Note that this feature is
to be ascribed to the use of relativistic kinematics in the energy-conserving
$\delta$-function of eq.(\ref{PWIA:resp}). In the nonrelativistic
regime the width of the bump in the quasielastic response increases
linearly with $q$, while its height displays a $(1/q)$ behavior.

When relativistic kinematics is used, the $y$-scaling variable for infinite 
nuclear matter is defined as
\beq
y = -q + \sqrt{ (\omega - E_{min})^2 + 2m(\omega - E_{min}) }\ ,
\label{svariable}
\eeq
$E_{min}$ being the minimum energy needed to remove a nucleon from
the nuclear matter ground state. The corresponding 
scaling function reads
\beq
F(q,y) =  \frac{q}{\sqrt{m^2 + (y + q)^2}}\ S({\bf{q},\omega}).
\label{sfunction}
\eeq

\begin{figure}
\centerline{\epsfig{figure=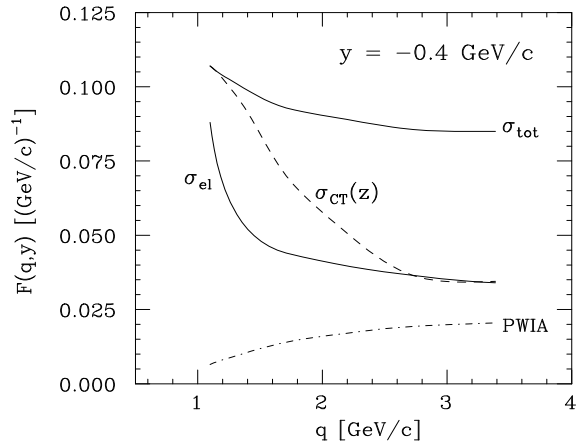,angle=090,width=8.50cm,height=6.1cm}}
\caption{
$q$-dependence of the nuclear matter scaling function at $y$ = - 0.4 GeV/c.
Dash-dot line: PWIA. Solid lines: FSI included using the 
parametrization of the NN scattering amplitude of 
refs.\protect\ci{NNAMP1,NNAMP2} with the total ($\sigma_{tot}$) or 
elastic ($\sigma_{el}$) cross section.  
 Dashed line: color transparency included according to the 
{\it quantum diffusion} model of ref.\protect\ci{CT}.
}
\label{fy}
\end{figure}

From the above equations, it can be readily seen that, if the 
folding function $F_q(\omega)$ and the shape of
the PWIA response become $q$-independent at large $q$, $F(q,y)$
obtained using $S({\bf q},\omega)$ of eq.(\ref{response})
is $q$-independent as well. Hence, in this case $F(q,y)$ {\it does} 
exhibit scaling but its value as $q \rightarrow \infty$ 
{\it does not} correspond to the PWIA limit, and cannot be simply 
related to the nuclear spectral function. In fig. \ref{fy} the PWIA
scaling function obtained using the spectral function of ref.\ci{bff},
(dot-dash line) is compared to the result of the calculation including
FSI, carried out with the parametrization of the NN scattering
amplitude of refs.\ci{NNAMP1,NNAMP2} (solid line labelled $\sigma_{tot}$).
Note that the PWIA $F(q,y)$ approaches $y$-scaling from below, whereas the
occurrence os FSI leads to a convergence from above. The solid line 
labelled $\sigma_{el}$ shows the results obtained neglecting
the contribution of inelastic processes, i.e. replacing the total cross section 
with the elastic cross section in the definition of the NN scattering amplitude
(see eq.(\ref{optical})). The different behavior of the two solid lines 
reflects
the fact that while $\sigma_{tot}$ changes by less than 10~\% over the range 
1~$< q <$~4 GeV/c, the ratio $\sigma_{el}/\sigma_{tot}$ drops from $\sim$ 95 \% 
to $\sim$ 30 \%. 

As pointed out in ref.\ci{gangofsix}, using the free-space amplitude to 
describe 
NN scattering in the nuclear medium may be questionable. Pauli blocking
and dispersive corrections are known to be important at moderate energies \ci{papi}.
However, their effects on the calculated nuclear response have been found to be small
in the kinematics of the SLAC data, corresponding to $q \sim$ 2 GeV/c, and
decrease as $q$ increases \ci{bho}. Corrections to the
amplitude associated with the extrapolation to off-shell energies are also
expected to be small at $q >$ 2 GeV/c \ci{bl}.

Modifications of the free-space NN cross section may also originate from the 
internal structure of the nucleon. It has been suggested \ci{CT1,CT2} that 
elastic scattering on a nucleon at high momentum transfer can only occur 
if the nucleon is found in the Fock state having the lowest number of constituents, 
so that the momentum can be most effectively shared among them. 
This state is very compact, its size being proportional to 1/$Q$, and therefore 
interacts weakly with the nuclear medium. Within this picture a nucleon, 
 after absorbing a large momentum $q$, travels through nuclear matter experiencing 
very little attenuation, i.e. exhibits {\it color transparency} (CT), and evolves 
back to its standard configuration with a characteristic timescale. 

The occurrence of CT is relevant to the analysis of inclusive electron-nucleus 
scattering at $y <$ 0, where elastic scattering is the dominant reaction 
mechanism, since it leads to a significant quenching of FSI. 
This effect has been quantitatively investigated in refs.\ci{gangofsix,bp,LDAnp}
using a specific model  developed by Farrar {\it et al.}\ci{CT}, referred to as 
{\it quantum diffusion} model, to describe 
the time evolution of the NN cross section associated with the onset of CT. 

According to ref.\ci{CT} the free space NN cross section  
$\sigma_{tot}(q)$ is recovered after travelling a distance 
$L = 2q/\Delta m^2$, with $\Delta m^2$ = 0.7 GeV$^2$, whereas at distances 
$z<L$ the cross section is suppressed and can be written in the form
\beq
\sigma_{CT}(q,z < L )  =  \sigma_{tot}(q) \left[ 
\frac{z}{L} + \frac{ 9\ \langle k_t^2 \rangle}{Q^2} 
\left( 1 - \frac{z}{L} \right) \right]\ ,
\label{sigma:CT}
\eeq
where ${\langle k_t^2 \rangle}^{1/2} \sim 0.35$ GeV/c. 

Substitution of the cross section of eq.(\ref{sigma:CT}) 
into the NN scattering amplitude obviously affects 
the momentum transfer dependence of FSI
effects, since the value of the ratio $\sigma_{CT}(q,z)/\sigma_{tot}(q)$ 
is reduced by a factor $\sim$(1/$Q^2$) at $z=0$ and evolves linearly
towards unity within a distance that grows linearly with $q$.

The results of the calculations of refs.\ci{gangofsix,bp,LDAnp} show that 
inclusion of CT effects according to the model of ref.\ci{CT}, involving no adjustable
parameters, greatly improves the agreement
between theory and data, allowing for a satisfactory description of the low 
energy loss tail of the nuclear inclusive 
cross sections, corresponding to large negative values of $y$, 
for $Q^2>1.5$~(GeV/c)$^2$.
 
The dashed line in fig. \ref{fy} shows that CT effects on the nuclear matter
$F(q,y)$ are large at $y=-0.4$ GeV/c and $q >$ 1.5 GeV/c. The scaling 
function obtained using the CT cross section of eq.(\ref{sigma:CT}) displays a 
distinct $q$-dependence, featuring a steep fall in the range 1 $< q <$ 3 GeV/c, 
followed by a plateau lying above the PWIA limit. At $q \sim$ 3 GeV/c the 
dashed line and the solid line labelled $\sigma_{el}$ come very close to
each other. This behavior has to be ascribed to the fact that inclusive
scattering at large $q$ is mostly sensitive to FSI taking place at distances 
$z \gtrsim$ $r_c$, $r_c \sim$ 0.5 fm being the radius of the repulsive
core of the NN interaction \cite{bho,bl}. In fact, using eq.(\ref{sigma:CT})
to model the $z$ dependence of $\sigma_{CT}$ one finds
$\sigma_{CT}(q,z \sim r_c) \sim \sigma_{el}$ at $q \sim$ 3 GeV/c. 

The analysis described in the present paper shows
that, when the rescattering processes are described in terms of free-space
NN scattering amplitude, the dominance of inelastic channels
 gives rise to sizeable FSI effects that
show no appreciable $q$-dependence and persist up to very large values of $q$.
As a consequence, the scaling behaviour of $F(q,y)$ cannot 
be interpreted as a signature of the onset of the PWIA regime.
A similar result was reached in ref.\ci{Negele}, 
 in the context of a study of $y$-scaling in the nonrelativistic hard-sphere 
Bose gas. 

Inclusion of CT effects leads to a sizeable suppression of FSI effects in the range 
1~$< q <$~4 GeV/c and to the appearance of a strong $q$-dependence. 

The results 
of fig. \ref{fy} suggest that the analysis of the scaling behavior of the TJNAF 
data may give a clue to the issue of the possible manifestation of CT in 
inclusive processes. However, it has to be kept in mind that extracting the 
experimental scaling function from the measured cross sections requires 
approximations and assumptions in the treatment of the electron-nucleon vertex.
Hence, a comprehensive quantitative comparison between theory and data 
at the level of the nuclear cross sections has to be regarded as a prerequisite 
to the $y$-scaling analysis.

The author is deeply indebted to V.R. Pandharipande and I. Sick
for many illuminating discussions. The hospitality of the TJNAF Theory Group 
 is also gratefully acknowledged.





\begin{references}

\bi{cg}
W. Czyz and K. Gottfried, Ann. Phys. (N.Y.) {\bf 45}, 47 (1963).

\bi{West}
G.B. West, Phys. Rep. {\bf 18}, 263 (1975).

\bi{3hescal}
I. Sick, D. Day and J.S. McCarthy, Phys. Rev. Lett. {\bf 45}, 871 (1980).

\bi{2hscal}
P. Bosted, R.G. Arnold, S. Rock and Z. Szalata, Phys. Rev. Lett.
{\bf49}, 1380 (1982).

\bi{NE3}
D. Day {\it et al.}, Phys. Rev. Lett. {\bf 59}, 427 (1987).

\bi{TJNAF}
J. Arrington {\it et al.},  Phys. Rev. Lett. {\bf 82}, 2056 (1999).

\bi{super}
T.W. Donnelly and I. Sick, Phys. Rev. Lett. {\bf 82}, 3212 (1999).


\bi{gangofsix}
O. Benhar, A. Fabrocini, S. Fantoni, G.A. Miller, V.R. Pandharipande
and I. Sick, Phys. Rev. C {\bf 44}, 2328 (1991).

\bi{bp}
O. Benhar and V.R. Pandharipande, Phys. Rev. C {\bf 47}, 2218 (1993).

\bi{LDAnp}
O. Benhar, A. Fabrocini, S. Fantoni and I. Sick, Nucl. Phys.
{\bf A579}, 493 (1994).

\bi{bff}
O. Benhar, A. Fabrocini and S. Fantoni, Nucl. Phys. {\bf A505}, 267 (1989).

\bi{butler}
M.N. Butler and S.E. Koonin, Phys. Lett. {\bf B 205}, 123 (1988).

\bi{bho}
O. Benhar, A. Fabrocini, S. Fantoni, V.R. Pandharipande, S.C. Pieper
and I. Sick, Phys. Lett. {\bf B 359}, 8 (1995).

\bi{NN:data}
A. Baldini, {\it et al.}, in {\it Total Cross Sections for Reactions of
High Energy Particles}. Landolt-B\"ornstein, New Series, Vol. I/12b, 
 edited by H. Schopper (Springer-Verlag, Berlin, 1987).

\bi{NNAMP1}
A.V. Dobrovolski, {\it et al.}, Nucl. Phys. {\bf B 214}, 1 (1983).

\bi{NNAMP2}
B.H. Silverman, {\it et al.}, Nucl. Phys. {\bf A 499}, 763 (1989).

\bi{papi}
V.R. Pandharipande and S.C. Pieper, Phys. Rev. C {\bf 45}, 791 (1992).

\bi{bl}
O. Benhar and S. Liuti, Phys. Lett. {\bf B389}, 649 (1996).

\bi{CT1}
S.J. Brodsky, in {\it Proceedings of the Thirteenth International Symposium
on Multiparticle Dynamics}, edited by E.W. Kittel, W. Metzger and A. Stergion 
(World Scientific, Singapore, 1982).

\bi{CT2}
A. Mueller, in {\it Proceedings of the Seventh Rencontre de Moriond}, 
edited by J.Tran Thanh Van (Editions Frontieres, Gif-sur-Yvette, 1982).

\bi{CT}
G.R. Farrar, H. Liu, L.L. Frankfurt and M.I. Strickman, Phys. Rev. Lett.
{\bf 61}, 686 (1988).

\bi{Negele}
J.J. Weinstein and J.W. Negele, Phys. Rev. Lett. {\bf 49}, 1016 (1982).

\end{references}
\end{document}